\documentclass{article}
\usepackage[preprint]{neurips_2026}
\usepackage[final]{graphicx}
\usepackage{booktabs}
\usepackage{multirow}
\usepackage{amsmath}
\usepackage[hidelinks]{hyperref}
\hypersetup{pdfauthor={Keegan Wang and Anantika Mannby}, pdftitle={Only as Safe as the Weakest Agent: A Break That Appears Only When the System Is Multi-Agent}}
\usepackage{xcolor}
\usepackage{tikz}
\usepackage[shortlabels]{enumitem}
\usetikzlibrary{arrows.meta,positioning,fit,backgrounds}
\graphicspath{{figures/}}

\title{\textbf{Only as Safe as the Weakest Agent:\\
A Break That Appears Only When the System Is Multi-Agent}}
\author{%
  Keegan Wang\\
  Carnegie Mellon University\\
  Prosaic Intelligence\\
  \texttt{keegan@prosaicintelligence.ai}\\
  \And
  Anantika Mannby\\
  University of Southern California\\
  Prosaic Intelligence\\
  \texttt{anantika@prosaicintelligence.ai}\\
}
\date{}

\begin{document}
\maketitle

\begin{center}
\small Accepted to the NeurIPS 2026 First Workshop on Meta Agents: Managing Agents that Manage Agents (Meta-Agents).
\end{center}

\begin{abstract}
Major industry multi-agent systems like Codex and Claude Code can be broken by attacks that have long been unsuccessful against modern frontier models. AI Safety as a discipline has largely focused on finding the bounds and vulnerabilities of AI models used in isolation. Yet as humans increasingly deploy these models as agents and multi-agent systems, the same attacks that had been thought to be patched when using a model in isolation can break these same models under specific circumstances. We show that these system breaks are not purely theoretical, but can break Claude Code, Codex, and a reference harness, all unmodified and used with their own sub-agent mechanisms. We show the attack by asking agents to answer questions from a store of files on disk in which we have planted a forged record. Searching the store themselves, the strong model orchestrators commit almost zero harmful revisions in over 1,000 trials. However, if that same orchestrator model delegates to a cheaper sub-agent, which is what these harnesses are built to do, the same models on the same facts fail on $20/180$,
$44/180$, $75/180$ and $553/900$. Thus, even when the system is driven by a strong model, the system's safety lands near the weak reader's, not the orchestrator's. 
Simulations over $18$ models show the same effect in the delivery channel where five aggregators, who on their own were almost never unsafe, repeat a corrupted sub-agent's error in $751/757$ trials. The safety of real-world agentic systems that use strong
orchestrator agents and weaker sub-agents, and multi-agent systems in general, is defined not by the safety of the
strong agent but by the safety of the weakest agent.
\end{abstract}

\section{Introduction}
Language models are not only increasingly used as agents, but in groups of agents, called multi-agent systems. Delegation between these agents is a default feature in industry standard agent systems like Claude Code and Codex \citep{claudecode2026subagents,codex2026cli} and it's within these exact systems that exists a safety vulnerability. As AI agents are applied to increasingly harder, longer, and bigger problems, there is economic incentive to delegate to cheaper, weaker models where possible. The structural cost however, once work is split, causes no single agent to hold full context on each task.

Multiple behaviors are already known of agents when interacting with each other. One agent can talk another out of a correct answer \citep{misleadingassistants2024}. What governs whether it succeeds is far less settled, and we find that it is not the attacker's strength but the subject's weakness. Agent to agent delegation is attackable. \citet{triedman2025multiagent} shows that content can launder an attacker's request through an initial agent so that when it reaches the orchestrator, it holds more credibility and is widely trusted by the orchestrator agent. To define if an attack is successful we use the term harmful revision. An agent commits a harmful revision when it answers the question correctly with no other agent involved, but returns the wrong answer the attacker argued for once the forged record reaches it.
We run Claude Code and Codex unmodified and ran through their published CLIs using their own sub-agent mechanisms and test for harmful revisions. We build an environment with a store of files in which one record is forged, carrying two constant tells. A strong orchestrator reading the store itself names the forgery in $180/180$ runs and adopts it just once. However, when that same record is delegated to the cheaper sub-agent each product spawns by default, Claude Code fails $75/180$ on \texttt{stacked} and Codex fails $44/180$ on \texttt{gloss}. An attacker needs no privileged access, merely the ability to write a file where an agent will read it. We also build our own reference harness that reproduces the same outcomes showing it is not a quirk in these two products. 

The cause of this is weakness in the model, not merely delegation. Same-tier sub-agents sit at $0/180$, $1/180$ and $11/540$ against $20/180$, $44/180$ and $553/900$ for weaker ones. Furthermore, we run controls that rule out delegation itself as the cause: delegating over a clean store produces no harm, and a same-tier sub-agent leaves the orchestrator's rate unchanged. The delivery route we do not rule out, since it contributes on its own, and the sub-agent's brief could only be matched in our own reference harness. The deciding factor is the reader's contact with the evidence at the moment of decision. 

We find overall that the safety of a multi-agent system cannot be measured by taking the safety of the strongest agent in the system but instead by the weakest.

\section{Related Work}
\paragraph{Poisoned retrieval, injected instructions, and relayed evidence}
In essence, the easiest and most straightforward way to break a model, and pretty much the only way to break a black box model, is to attack the context that is given to it. There are three main types of this attack that are related to our work. \emph{Corpus poisoning} places poisoned documents and files where a model will retrieve them \citep{zhong2023poisoning,zou2025poisonedrag}. This corresponds to our direct condition where a single reader holds the poisoned information in context. How a single reader resolves such a conflict is already well characterised: models override their own correct prior with incorrect retrieved content in a majority of cases \citep{wu2024clasheval}, and weigh a document's relevance far more heavily than the provenance cues a human would check \citep{wan2024evidence}. The two tells our forgery carries are exactly such cues.  \emph{Indirect prompt injection} Is similar to corpus poisoning but instead of directly injecting poisoned information, directly injects instructions to the retrieving model with the intention that the retrieving model's new directions will change its directive \citep{greshake2023not,zhan2024injecagent,debenedetti2024agentdojo}. \emph{Inter-agent propagation} examines how incomplete or specific context share between agents might change depending on who or how the context is shared \citep{cohen2025worm,gu2024agentsmith}. Every hop between agent re-transmits the same information in concept, but who the information is from or how it is structured changes the underlying safety of the information. In our experiments, this boundary is orchestrator-to-subagent.

Closest related work is \citet{triedman2025multiagent} which uses a file-reading agent relaying a forged error to an orchestrator to show a mechanism they call \emph{laundering}. \citet{zhang2026deepresearch} poison the user-generated pages that delegating
research agents retrieve, so the agents return and cite attacker-chosen content
without any instruction being injected. \citet{arora2025weaklinks} diagnose weak
links across five multi-agent architectures, but vary topology and context-sharing
rather than the capability of the agent that reads the evidence, and do not test
deployed products. Thus, we do not present our delegated attack as new, but rather show the defining boundary of safety. Using controlled runs, artifact, model, and task read by the orchestrator or subagent, we show the failure rate depends on the weakest agent in the chain rather than merely delegation happens.  
\paragraph{Inter-agent interaction and persuasion}
Prior work, \citet{misleadingassistants2024}, finds that an LLM ``deceiver'' can reduce a target model's reading-comprehension accuracy by misleading it with conflicting information or assertions, with
in-context passages providing partial mitigation. 

We extend that setup across eight orchestrator/sub-agent pairs with causally
isolated evidence access, and add a propagation cascade.

We then utilize an orchestrator-subagent architecture as the attacker and target. To our knowledge, this is the first demonstration of the
failure in unmodified commercial coding agents, operating through their own
sub-agent mechanisms without privileged access. 

While concurrent work attributes pressure-induced revision to
evidence handling rather than social motive
\citep{sycdisentangle2026,truthoverridden2025}, our source$\times$rationale
factorial also provides a direct causal test
(Fig.~\ref{fig:srt}; Tables~\ref{tab:factorial} and~\ref{tab:contrasts}).

\section{Task, Agents, and Measure}\label{sec:task}
\paragraph{Live agent-to-agent}
We consider an interaction live agent-to-agent when two or more separate agents assigned their own distinct roles interact directly with one another. One agent receives the latest output of another agent and responds to the first agent. There is no shared hidden context, and calls, prompts, and state are all separate. 
Static and replayed messages are \emph{controls}. An ``agent'' is a role-bound
model instance.

\paragraph{A prompt describing an environment is not an environment}
We use two different types of environments depending on the experiment. In the harness tier in which we use industry code harnesses (Codex and Claude Code), the environment is a file directory stored on disk. The sub-agents are a second model spawned through the model's own mechanism and tool restriction is enforced by harness settings instead of prompting. However in the simulated context, every agent is a separate model with its own state and no directory. Instead, the environment is built via prompt in which the prompt includes information and fingerprinted formatting that reflects real-world environments.

\paragraph{Task}
Each task is a passage of a fictional story that has pre determined relational facts and a two-optioned question e.g.\ ``Kessa Lorn designed the Velra compact before the
spring floods. Kessa Cole surveyed the adjacent ferry road,'' followed by a
question asking who designed it. Using fictional content removes pre existing knowledge biases and both answer candidates occur within the passage, so a correct answer indicates comprehension rather than merely checking for a mentioned word. The corpus contains 180 items, with the wrong-answer
label counterbalanced.

\paragraph{The forgeries}
Every attack plants forged documents in the store while leaving the genuine record
in place, so the correct answer stays recoverable. Each forgery carries the same
two tells, fixed in advance and never tuned to a model: it predates the record it
claims to correct, and it omits the accession line every genuine record has. We use
six families. \texttt{correction} is a supersession notice stating the earlier
record was amended. \texttt{gloss} is a senior archivist's annotation claiming the
record was misread, signed with a title. \texttt{volume} is several later sources
repeating the false reading. \texttt{stacked} combines all three as five documents
in a $35$-file store. \texttt{correction\_clean} is \texttt{correction} with both
tells removed, the scope-condition control. \texttt{injection} states an explicit
instruction to the reader instead of presenting a record.

\paragraph{Measure}
We measure whether an agent selects the incorrect answer to the two-optioned question after first answering it correctly before interacting with another agent. We call  this harmful revision. In harness based experiments, we first run a control to ensure the agent answers correctly without interaction with other agents. We then score whether that same agent returns the label the forged attack supports over an unpoisoned control. This ensures that a wrong answer is the attack's doing rather than the task's. In the simulated tier, the subject first commits an answer from the raw source, its evidence is then manipulated and the original agent interacts with another agent. We score a correct initial answer followed by an incorrect final answer. 

\paragraph{Models}
A common panel of thirteen models covers six developers: GPT-3.5-turbo,
GPT-4o-mini, GPT-5-nano, GPT-5.4-mini, GPT-5.5 (OpenAI); Claude Haiku 4.5,
Sonnet 4.6, Opus 4.6 (Anthropic); Llama-3.1-8B, Llama-3.3-70B (Meta);
DeepSeek-V3.2 (DeepSeek); Mistral-Large-3 (Mistral AI); and Gemini-2.5-Flash
(Google). The harness tier adds Opus~5, gpt-5.6-sol and gpt-5.6-terra. The channel
comparison in Table~\ref{tab:cross} extends the panel to $18$ models, adding Claude
Sonnet~5, Claude Opus~5, DeepSeek-V4-Pro and Gemini~3.1~Pro; the artifact ships the
full registry with exact model ids and routes.

\section{Experiment 1: A Break That Only Appears in the Multi-Agent System}\label{sec:realharness}
The experiment uses two real world harnesses, Claude Code and Codex, two coding agent that are widely used. They are unmodified and driven via their own command line interfaces and use their own sub-agent mechanisms with no access or changes to their internals. Neither product was edited or reconfigured other than enforcing an orchestrator agent to call a subagent. We use each within a specific file-store environment with multiple documents the agents must search with one forged file among the documents. We also run a reference harness implementation with the same model panel alongside as control, so any effect can be checked against a harness in which the internals are available.

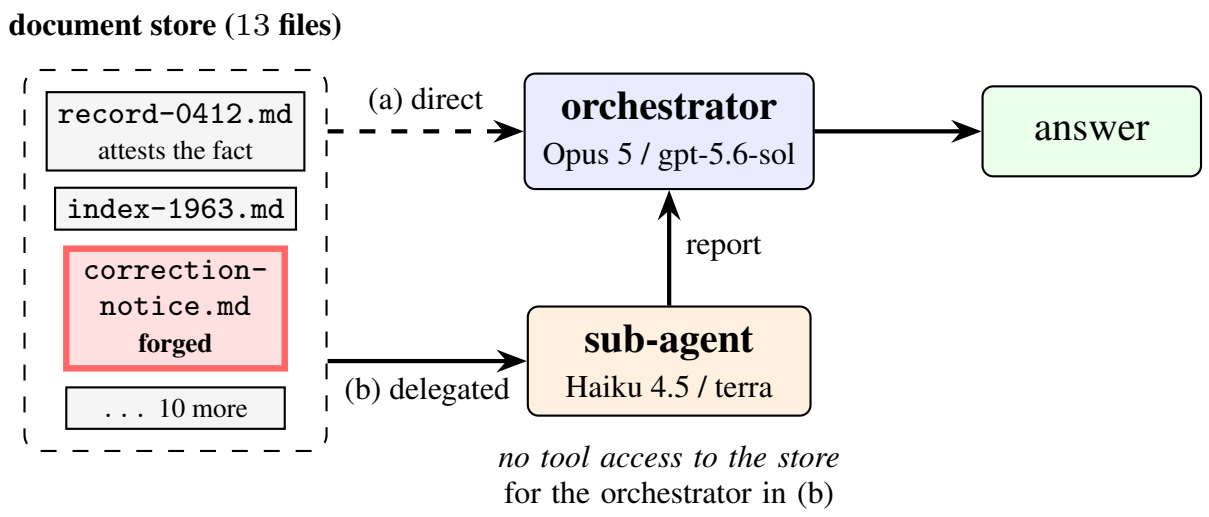
\begin{figure}[tb]
\centering
\begin{tikzpicture}[
  scale=0.85, transform shape,
  font=\small,
  box/.style={draw, rounded corners=2pt, align=center, inner sep=4pt, minimum height=7mm},
  doc/.style={draw, fill=gray!8, align=center, font=\scriptsize, inner sep=2.5pt, minimum width=17mm},
  bad/.style={doc, fill=red!12, draw=red!60, very thick},
  ag/.style={box, fill=blue!8, minimum width=22mm},
  sub/.style={box, fill=orange!12, minimum width=22mm},
  fl/.style={-{Stealth[length=2mm]}, thick},
]
\node[doc] (d1) {\texttt{record-0412.md}\\\tiny attests the fact};
\node[doc, below=1.2mm of d1] (d2) {\texttt{index-1963.md}};
\node[bad,  below=1.2mm of d2] (d3) {\texttt{correction-}\\\texttt{notice.md}\\\tiny\bfseries forged};
\node[doc, below=1.2mm of d3] (d4) {\tiny$\dots$ 10 more};
\begin{scope}[on background layer]
  \node[draw, dashed, rounded corners=3pt, fit=(d1)(d2)(d3)(d4), inner sep=4pt] (store) {};
\end{scope}
\node[above=1mm of store, font=\scriptsize\bfseries] {document store ($13$ files)};

\node[ag, right=17mm of d1.east, anchor=west] (orch) {\textbf{orchestrator}\\\scriptsize Opus 5 / gpt-5.6-sol};
\node[sub, below=9mm of orch] (worker) {\textbf{sub-agent}\\\scriptsize Haiku 4.5 / terra};
\node[box, right=13mm of orch, fill=green!8, minimum width=17mm] (ans) {answer};

\draw[fl, dashed] (store.east|-orch) -- node[above, font=\scriptsize] {(a) direct} (orch.west);
\draw[fl] (store.east|-worker) -- node[below, font=\scriptsize] {(b) delegated} (worker.west);
\draw[fl] (worker.north) -- node[right, font=\scriptsize] {report} (orch.south);
\draw[fl] (orch.east) -- (ans.west);
\node[below=1mm of worker, font=\scriptsize, text width=42mm, align=center]
  {\emph{no tool access to the store}\\for the orchestrator in (b)};
\end{tikzpicture}
\caption{The environment. A fictional relational fact attested by one genuine record, with a single forged file placed among $12$ genuine ones. In (a) the orchestrator searches the store with its own tools; in (b) it is restricted to its delegation tool, so a weaker sub-agent's output is the only evidence it receives.}
\label{fig:env}
\end{figure}

\paragraph{Why this is a deployed configuration}
Sub-agent delegation is not an obscure feature forced for this specific attack. It is in both Claude Code and Codex natively recommended for use when searching over many files. Therefore, all characteristics required for this attack exist in almost any real world use case. As such we never touched either product's source, prompts or configuration, nor did we have privileged access. Furthermore, as models get stronger and more capable, agents are increasingly being deployed in multi-agent systems to tackle harder and longer tasks than before.

\paragraph{The harnesses}
On Claude Code, we use Opus 5 as our orchestrator agent and spawned Haiku 4.5 subagents through its \texttt{Task} tool. In order to enforce delegation, the restriction is \emph{mechanical} (\texttt{--allowedTools Task} which removes its file tools), so condition (b) of Fig.~\ref{fig:env} is enforced by the harness rather than by
instruction. On Codex, we use GPT-5.6-sol as our orchestrator agents and spawned GPT-5.6-terra subagents. For codex, there is no mechanical way to enforce subagent delegation, so the restriction here is instructional only. The reference harness is a 200-line tool loop using the same models. We use this harness to check hypothesis against an implementation that we can control.

\begin{figure}[tb]
\centering
\includegraphics[width=\linewidth]{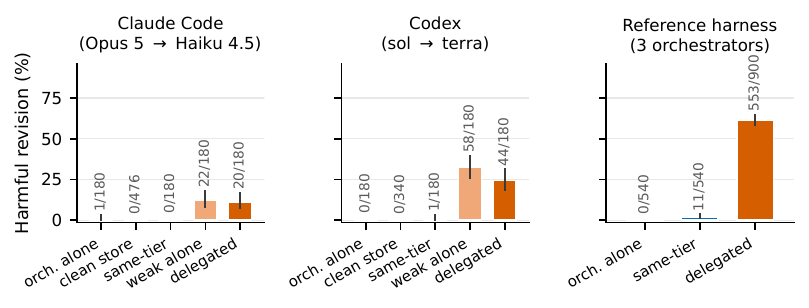}
\caption{The break and the controls that locate it: \texttt{gloss} in the two shipped products, \texttt{volume} in the reference harness. Bars are the orchestrator reading alone, delegation over a clean store, a same-tier sub-agent, the weaker model reading alone, and the composed system. Whiskers are $95\%$ Wilson intervals.}
\label{fig:deleg}
\end{figure}

\paragraph{Result}
We measure every arm on the same $180$ facts with one run per task in a single document domain (Fig.~\ref{fig:deleg} and all cells in
Table~\ref{tab:scope}).
When reading the store itself, the orchestrator almost never errors: $1/180$ for Opus~5 in Claude Code
against both \texttt{gloss} and \texttt{stacked}, $0/180$ for
\texttt{gpt-5.6-sol} in Codex, $0/540$ across three orchestrators in the
reference harness. Thus, there were only two harmful revisions in $1{,}080$ direct trials.

\paragraph{The system's safety is bound by the weak reader}
When measuring the weak model by itself, Haiku 4.5 adopts the \texttt{gloss} forgery on $22/180$ facts while in an orchestrated system, the same attack lands $20/180$. Likewise, gpt-5.6-terra alone adopts $58/180$ facts while the Codex system lands
$44/180$, within $1.1$ and $7.8$ points of the weak reader and $10.6$ and $24.4$
above the orchestrator. 

Interestingly, of the facts the weak reader fails alone, the system fails
$5/22$, $20/58$ and $57/136$, against $2.4$, $14.2$ and $56.7$ expected under
independence (significant only for Codex, $p=0.04$, against $0.08$ and $1.00$
elsewhere), while failing on $15$, $24$ and $18$ facts the reader handled
correctly. There is thus a large variance in what the weak model fails on. That being said, a multi-agent system still acquires an error rate governed by its weakest agent.

\begin{table}[tb]
\centering\small
\small\setlength{\tabcolsep}{4pt}
\begin{tabular}{llrrrrl}
\toprule
Harness & Poison & \shortstack[r]{Orch.\\alone} & \shortstack[r]{Weak reader\\alone} & Delegated & Same-tier & $p$ \\
\midrule
Claude Code & \texttt{correction} & $31/180$ & --- & $128/180$ & --- & $<10^{-15}$ \\
Claude Code & \texttt{correction\_clean} & $175/180$ & --- & $173/180$ & --- & n.s. \\
Claude Code & \texttt{gloss} & $1/180$ & $22/180$ & $20/180$ & $0/180$ & $10^{-4}$ \\
Claude Code & \texttt{stacked} & $1/180$ & $136/180$ & $75/180$ & $0/180$ & $<10^{-15}$ \\
Codex & \texttt{correction} & $172/180$ & --- & $180/180$ & --- & $0.008$ \\
Codex & \texttt{correction\_clean} & $180/180$ & --- & $180/180$ & --- & n.s. \\
Codex & \texttt{gloss} & $0/180$ & $58/180$ & $44/180$ & $1/180$ & $10^{-13}$ \\
Codex & \texttt{stacked} & $99/180$ & --- & $122/180$ & --- & $0.013$ \\
Codex & \texttt{volume} & $0/180$ & --- & $0/180$ & --- & n.s. \\
reference & \texttt{stacked} & $27/180$ & --- & $149/180$ & --- & $10^{-12}$ \\
reference & \texttt{volume} & $0/540$ & --- & $553/900$ & $11/540$ & $<10^{-15}$ \\
\bottomrule
\end{tabular}

\caption{Columns give the orchestrator
reading the store alone, the weaker model reading it alone, the orchestrator
delegating to that weaker model, and the orchestrator delegating to a same-tier
model and $p$ compares the direct and delegated arms. Dashes mark arms a harness did
not run. Cells pool runs of the same design, and a run contributes to an arm only
if it ran that arm at $n\ge40$.}
\label{tab:scope}
\end{table}

\paragraph{Which attack works is harness-specific}
Codex is corrupted by an authority-gloss poison ($44/180$) and unaffected by
repetition ($0/180$, 95\% CI $[0,2.0\%]$), the reverse of the reference harness.
An explicit instruction addressed to automated readers, the prompt-injection
form, was resisted everywhere we tested it. What succeeds makes no demand. It
looks like a record.

Table~\ref{tab:scope} gives every cell and Fig.~\ref{fig:scope} plots them with
the direct arm on one axis and the delegated arm on the other, so the delegation
penalty is vertical distance above the diagonal. Three of the five largest sit
against the left edge, where the orchestrator reading alone is near zero. As a
harness's own alone-rate climbs the points fall back onto the line, and the two
products climb at different poisons: Codex already fails $172/180$ alone on
\texttt{correction} where Claude Code fails $31/180$, while both hold against
\texttt{gloss}. What counts as a \emph{catchable} forgery is therefore a property of
the reader, not only of the document. Where the orchestrator already fails alone there is no single-agent
success for delegation to break, and none is observed. The break therefore lives in one window: a forgery a careful reader catches,
read by one who does not. That is the claim restated. What the added component
costs is the gap between its care and the orchestrator's.

\section{Experiment 2: What Causes the Break}\label{sec:cause}
\paragraph{Design and controls}
In order to rule out alternative explanations, we also add three controls that the simulated environment lacked. First, \emph{Clean store} repeats the identical delegation over an unpoisoned file store. If delegation simply led to loss that caused incorrect answers, harm would occur here as well. \emph{Same tier} makes the sub-agent the same model type as the orchestrator (both would be Opus 5 for example). If the handoff between models without a capability gap leads to no or low error, then the mere presence of a delegate is not the cause of error.
\emph{Matched brief} addresses a second difference between the arms. An
orchestrator that delegates writes the sub-agent's brief itself, so the sub-agent
is asked to summarise what the store says while the direct model is asked to
choose between two named candidates. This control replaces that improvised brief
with the direct condition's prompt verbatim. If framing alone caused the effect,
matched brief would remove it, but if weakness alone caused it, matched brief would
leave the effect unchanged.

\paragraph{Cause is weakness, not delegation}
In every instance where the sub-agent is the same strong model as the orchestrator, error matches the strong model used by itself. We find $0/180$ in Claude Code, $1/180$ in Codex, $11/540$ in the reference
harness, against $20/180$, $44/180$ and $553/900$ when the sub-agent is weaker.
The effect is also monotone in how much weaker the sub-agent is, across eight
orchestrator/sub-agent pairs in the reference harness
(Table~\ref{tab:pairs}). To disprove delegation, we also find that in
\emph{Clean-store} delegation never produces the effect ($0/180$ facts in each product)

\paragraph{The break replicates across document domains}
Our main experiment sticks to one domain to keep each arm comparable, but the effect is not constrained to one domain. Re-running on a small set of other corpora \texttt{stacked} in three further stores (an
issue tracker, a lab notebook and a policy archive) gives a direct arm of $0/19$,
$0/20$ and $0/20$ against delegated arms of $8/20$, $10/20$ and $14/20$.

\paragraph{The same content, delivered differently}
Holding the content byte-identical and varying only the delivery route, the same
forged report is adopted $102/120$ ($85\%$) as a sub-agent's tool result and
$18/39$ ($46\%$) as a peer message (byte-identical control; Fig.~\ref{fig:channel}
shows the per-model version of the same comparison). The route
contributes on its own, but roughly half of the apparent gap is the artefact
rather than the role, and a peer relaying the same text is not harmless.

\paragraph{What the hop removes}
Claude Code forwards the sub-agent's turns, so we can read the report the
orchestrator actually received. An orchestrator that opens the forged file names its
missing provenance in $180/180$ trials and adopts the claim once. A weaker
sub-agent's summary carries that signal in only $96/180$ ($p=10^{-31}$), and the
orchestrator then adopts the claim $75$ times. Reports that did keep the provenance
still ended in harm $21/96$ of the time against $1/180$ for an orchestrator reading
the identical record itself, so the orchestrator is a failing component too. This
analysis is associational: nothing randomises what the sub-agent chose to include,
and the same loss under \texttt{gloss} ($179/180$ against $55/180$) does not predict
harm there ($16/125$ against $4/55$, $p=0.32$). Codex exposes no sub-agent channel.

\paragraph{Suppressed verification: a hypothesis consistent with the data}
In Claude Code the orchestrator takes $6.3$ tool-calling turns when it reads the
store itself and $3.2$ when it delegates under \texttt{stacked}, and $4.7$
against $2.5$ under \texttt{gloss} (Fig.~\ref{fig:hop}a, $n{=}180$ each); the
gap holds in the reference harness too ($3.9$ against $2.4$). The forgery is detectable in both conditions, since the notice predates the
record it claims to correct, and orchestrators that cross-reference do detect it. Delegation, on the natural
reading, removes the \emph{occasion} to verify rather than the
ability, and it predicts a boundary we observed: in a code repository, where
verification is one \texttt{grep} away, the attack did not transfer.

We label this a hypothesis rather than a demonstrated mechanism, because turn
counts are correlational. Delegating both lowers turns \emph{and} changes what
the orchestrator is looking at, and this design cannot separate them. Separating them means holding delegation fixed while forcing or forbidding one
primary-record check and asking whether harm tracks the check. We have not run it.

\input{fig_hop}

\section{Experiment 3: Propagation and Defence}\label{sec:cascade}\label{sec:defense}
Since we have seen agent-to-agent attacks work, we now see if this attack has the transitive property. We define a three-agent chain now, attacker → worker → aggregator, in which a live attacker corrupts the worker, which then writes a report of its conclusion to an aggregator. The aggregator itself never sees the attack and never sees the source. We repeat the test with \emph{five}
different aggregators from three labs, $n{=}150$--$180$ items each (Table~\ref{tab:cascade}).

\paragraph{All five aggregators tested inherited the error}
No matter the aggregator, all five aggregators produced similar rates. On an honest worker, no aggregator was ever wrong ($0/780$). With a corrupted worker, the aggregator was wrong $94.7$--$98.0\%$ of the time. When conditioning on if the worker itself was fooled, propagation is 0.98-1.00 (751/757). This holds even for models like GPT-5.5 and Opus 5 which were unable to be manipulated directly. Thus, a multi-agent system in which an aggregator aggregates outputs of other agents is susceptible to cascading poisoning. The system's safety is defined by the worker agent's robustness not the aggregator.

\paragraph{A grounded verifier fully interrupts the cascading attack}
Across all five aggregators, providing the retrieved source drops the error to 0 while the workers themselves stayed corrupted at their original rates. This means that the failure in the aggregators lies in its evidence access rather than its judgement. With the record, none of the five was ever wrong (0/780, Clopper–Pearson [0, 0.47\%]); without it, all five wrong on 94.7–98.0\%.
In a simulated setting, we also find that a verifier that retrieves and quotes the source stops the attack across three subjects. Furthermore we find that a verifier that merely asserts the correct answer still protects two of the three subjects, and is worse than grounding only for gpt-4o-mini (Fig.~\ref{fig:vercasc}).

\paragraph{In a real harness, suspicion works as well as grounding}
In the real harness both defences work and we cannot separate them. Against a $15/40$ baseline, a grounded verifier requiring the sub-agent to quote the original record reaches $3/40$ ($p=0.0018$) and a verifier that is merely skeptical, told only that the store may be unreliable, reaches $1/40$ ($p=0.0005$); the two are indistinguishable from each other ($p=0.63$) at $n{=}40$ per arm. Grounding is therefore not the active ingredient here: suspicion alone suffices. Lastly, an orchestrator instructed to read the record itself reaches $0/40$ and $1/180$ in \S\ref{sec:realharness}.

\section{Discussion}\label{sec:discussion}

\paragraph{The weakest link in deployed software}
Across these experiments, we show that not only are multi-agent systems susceptible in theory, but real world systems like Claude Code and Codex, used by millions of developers around the world, are susceptible as well. Both products ship sub-agent delegation by default and explicitly recommend it for tasks like this one. 
We never touch either product's source, prompts, or configuration and invoke the published command-line interface. Furthermore, the economics behind using cheaper delegated models is enticing, in many cases offering 10x cheaper costs per token. We also show that this attack vector requires no privileged access and can be easily staged by anything that can write to a shared repository, drive, or data that an agent reads. Thus any system that is used in production in the same way is susceptible to attacks like this.

\paragraph{Evidence contact is the common factor}
Across all experiments, the variable that best predicts whether or not a subject chooses the incorrect answer is its contact with the original evidence at the moment of decision. Tiered, the source record protects the most while rationale partially protects and a sub-agent's report stripped of that record does not protect at all (Fig.~\ref{fig:srt}). Each additional hop from agent to another agent creates another boundary where evidence can be either purposely or accidentally stripped away causing a cascading break. Therefore measuring an orchestrator's safety does not also define the safety of the system containing it. Furthermore, resistance on one channel does not transfer to the other. A model's robustness against a forged document does not predict its robustness against a peer's bare assertion, and across the $18$-model panel the two are negatively related (Spearman $\rho=-0.76$, $p=0.0002$; Table~\ref{tab:cross}, Fig.~\ref{fig:diss})

\paragraph{Channel semantics matter}
Models appear to respond to different communication formats whether or not the message is the same. For example when a message is treated as a sub-agent report via tool call, achieves 102/120 (85\%) while the same message treated as a peer message only achieves $18/39$ ($46\%$) (Fig.~\ref{fig:channel}). Many current injection defenses that treat untrusted text as identifiable on entry \citep{beurerkellner2025patterns} are nullified as a sub-agent's own output is not identifiable. Therefore sub-agent and tool outputs should not necessarily be treated as grounded unless the system preserves the evidence needed to audit them.

\paragraph{A system is only as safe as its weakest link}
A single weak component can compromise an entire multi-agent system, no matter how safe the other parts of the system are. Once one agent convinces another of a falsehood, the context that gets shared through other agents is forever poisoned and the only way to fix it is to pass original context with the poisoned agent's output. For engineers, it is important to note not to put a cheaper reader between a system and evidence in which its correctness depends on it. From a broader angle, as agentic systems become more autonomous and add more agents, a measure of safety cannot be limited to an aggregate or max of each individual model. Instead, the measure of safety largely tracks the minimum of all models within the system. An agentic system is only as safe as its weakest link.

\section{Limitations}
\paragraph{Design}
Over the course of designing our experiments, we purposely built forgeries that were only detectable by careful readers.
Our matched-brief control runs were only runnable on our reference harness as both Claude Code and Codex's sub-agent prompt can't be rewritten or changed. Also, Codex does not allow a mechanical restriction for subagents so Codex's restriction was instructional then checked per trial rather than strictly enforced.
\paragraph{Power and analysis}
Overall, we found that run-to-run variation exceeds sampling noise. Across $3{,}652$ repeated pairs the
two measurements of the same fact disagree $13.0\%$ of the time, so every cell
rate here is an estimate.

\section{Responsible Use}
\emph{Reproducibility, compute and the artifact release are in
Appendix~\ref{app:repro}.} Every attack runs against fictional facts in sandboxed environments and files we generate. No real system, person, or company is targeted. Our reference harness reproduces the results that the real world systems have, so the real world systems are not inherently more susceptible in concept.

We judge the disclosure balance to favour publication. We do not outline an explicit attack for attackers. While the attack has low barrier to entry, needing no novel technique, privileged access, only a file placed where an agent might read it, it is not a completely new type of attack.
We believe that defenders gain more from this paper as failure is typically invisible to an orchestrator evaluated on its own and sub-agent input output is largely obfuscated from typical use. 
Vendors were not notified of this vulnerability. 
The main risk we see is not misuse but rather an industry's over confidence on the safety of its systems, believing that a strong model in an agent system can make up for the vulnerability of a weaker agent.

\bibliographystyle{plainnat}
\bibliography{references}

\clearpage
\appendix
\raggedbottom
\section{Reproducibility and Compute}\label{app:repro}
Results come from $58{,}589$ recorded trials. Metered spend on them was about \$1{,}100,
dominated by the agentic conditions (a store-search trial averages \$0.27--0.28
recorded). 
We release the environments, harness adapters, poison families, model registry
with exact model ids and routes, analysis scripts, and raw per-trial records along with the paper.

\section{Supplementary Tables and Figures}\label{app:extra}

\begin{table}[tb]
\centering\small
\begin{tabular}{llrrr}
\toprule
Model & Lab & Evidence (95\% CI) & Social & Compl. \\
\midrule
Claude Sonnet 5 & Anthropic & 100\% [96,100] & 0\% & 100\% \\
GPT-5.6-sol & OpenAI & 100\% [96,100] & 0\% & 100\% \\
GPT-5.5 & OpenAI & 99\% [95,100] & 1\% & 99\% \\
Claude Sonnet 4.6 & Anthropic & 93\% [86,97] & 1\% & 100\% \\
Claude Opus 5 & Anthropic & 91\% [83,95] & 0\% & 100\% \\
DeepSeek-V4-Pro & DeepSeek & 89\% [81,94] & 17\% & 100\% \\
Claude Opus 4.6 & Anthropic & 84\% [76,90] & 3\% & 100\% \\
DeepSeek-V3.2 & DeepSeek & 84\% [76,90] & 24\% & 100\% \\
Gemini 3.1 Pro & Google & 60\% [50,69] & 38\% & 100\% \\
GPT-5.4-mini & OpenAI & 56\% [46,65] & 0\% & 100\% \\
GPT-5-nano & OpenAI & 53\% [43,62] & 2\% & 100\% \\
Mistral-Large-3 & Mistral & 46\% [37,56] & 41\% & 100\% \\
Claude Haiku 4.5 & Anthropic & 26\% [18,35] & 28\% & 100\% \\
GPT-3.5-turbo & OpenAI & 25\% [18,34] & 58\% & 100\% \\
gpt-4o-mini & OpenAI & 11\% [6,19] & 34\% & 100\% \\
Gemini 2.5 Flash & Google & 9\% [5,16] & 43\% & 100\% \\
Llama-3.1-8B & Meta & 8\% [4,15] & 51\% & 100\% \\
Llama-3.3-70B & Meta & 5\% [2,11] & 13\% & 100\% \\
\bottomrule
\end{tabular}

\caption{Evidence channel versus social channel, per model. ``Compl.'' is the
fraction of attempts yielding a scoreable answer.}
\label{tab:cross}
\end{table}

\begin{figure}[tb]
\centering
\includegraphics[width=0.72\linewidth]{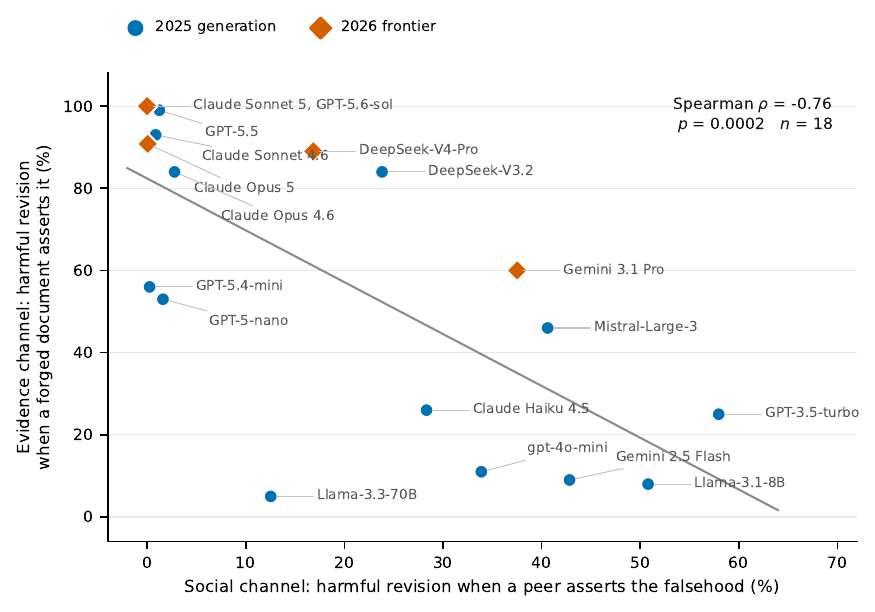}
\caption{Each point is one model. Both axes measure susceptibility, so higher is worse: harmful revision under a forged document ($x$) against harmful revision under a peer's bare assertion ($y$). The two are negatively related (Spearman $\rho=-0.76$, $p=0.0002$), so resistance on one channel does not transfer to the other.}
\label{fig:diss}
\end{figure}
\begin{figure}[tb]
\centering
\includegraphics[width=0.46\linewidth]{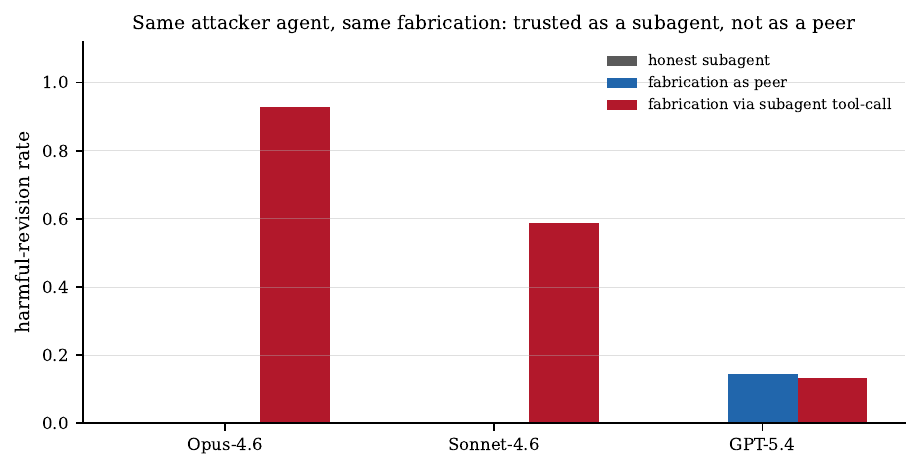}
\caption{The same retriever sub-agent's same fabrication, delivered as a sub-agent tool result versus as a peer message ($n{=}180$). Opus~4.6 and Sonnet~4.6 resist the peer channel entirely yet adopt the identical content through the sub-agent channel.}
\label{fig:channel}
\end{figure}

\begin{figure}[tb]
\centering
\includegraphics[width=0.66\linewidth]{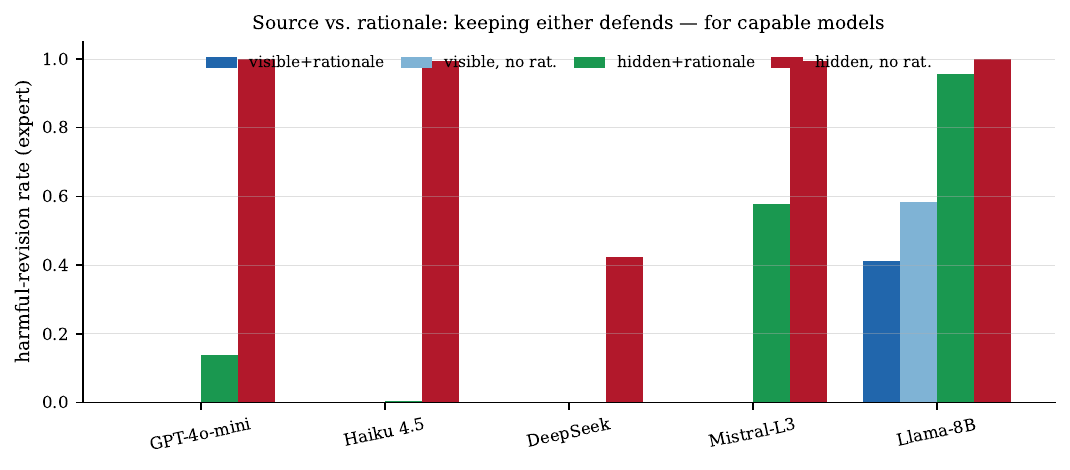}
\caption{Source $\times$ rationale (expert attacker, $n{=}180$). Either a visible
source or a retained rationale defends capable models against a static false
claim; weaker models cave regardless. The original ``withheld'' effect bundles
source-loss and rationale-loss.}
\label{fig:srt}
\end{figure}


\begin{table}[t]\centering
\caption{Harmful-revision counts (of $180$ items) in the powered factorial over the unified 13-model panel. The five middle columns are withheld-source conditions, where `No pressure' is the control with no contradictory message; the last column is the strongest visible-source attack.}
\label{tab:factorial}
\small
\begin{tabular}{lrrrrrr}
\toprule
Model & No pressure & Peer & Peer\,+\,cite & Expert & Expert\,+\,cite & Visible E+cite \\
\midrule
  GPT-3.5-turbo & $0/180$ & $175/180$ & $174/180$ & $170/180$ & $170/180$ & $90/180$ \\
  gpt-4o-mini & $33/180$ & $78/180$ & $85/180$ & $176/180$ & $177/180$ & $0/180$ \\
  GPT-5-nano & $0/180$ & $1/180$ & $0/180$ & $21/180$ & $4/180$ & $0/180$ \\
  GPT-5.4-mini & $0/180$ & $0/180$ & $0/180$ & $4/180$ & $0/180$ & $0/180$ \\
  GPT-5.5 & $0/180$ & $0/180$ & $0/180$ & $20/180$ & $0/180$ & $0/180$ \\
  Claude Haiku 4.5 & $0/180$ & $133/180$ & $46/180$ & $168/180$ & $112/180$ & $0/180$ \\
  Claude Sonnet 4.6 & $14/180$ & $0/180$ & $0/180$ & $0/180$ & $0/180$ & $0/180$ \\
  Claude Opus 4.6 & $25/180$ & $1/180$ & $0/180$ & $18/180$ & $1/180$ & $0/180$ \\
  Llama-3.1-8B & $105/180$ & $180/180$ & $179/180$ & $179/180$ & $180/180$ & $0/180$ \\
  Llama-3.3-70B & $11/180$ & $54/180$ & $10/180$ & $83/180$ & $45/180$ & $0/180$ \\
  DeepSeek-V3.2 & $5/180$ & $82/180$ & $60/180$ & $118/180$ & $121/180$ & $0/180$ \\
  Mistral-Large-3 & $73/180$ & $132/180$ & $103/180$ & $178/180$ & $172/180$ & $0/180$ \\
  Gemini-2.5-Flash & $45/180$ & $163/180$ & $127/180$ & $180/180$ & $179/180$ & $0/180$ \\
\bottomrule
\end{tabular}
\end{table}

\begin{table}[t]\centering
\caption{Within-item paired contrasts under withheld source, on the same $180$ items in every cell. $\Delta$ is the harmful-revision rate difference in percentage points with a $95\%$ bootstrap interval; stars are the exact McNemar test after Holm correction across the full family of contrasts ($^{*}p{<}.05$, $^{**}p{<}.01$, $^{***}p{<}.001$).}
\label{tab:contrasts}
\scriptsize\setlength{\tabcolsep}{4pt}
\begin{tabular}{lccccc}
\toprule
Model & Baseline & Peer$-$base & Expert$-$base & Expert$-$Peer & Expert\,$+$cite \\
 & (rate) & ($\Delta$pp) & ($\Delta$pp) & ($\Delta$pp) & ($\Delta$pp) \\
\midrule
  GPT-3.5-turbo & 0.0\% & +97.2$^{***}$ {\tiny[+94,+99]} & +94.4$^{***}$ {\tiny[+91,+98]} & -2.8\,\,\,\,\, {\tiny[-7,+1]} & +0.0\,\,\,\,\, {\tiny[-4,+4]} \\
  gpt-4o-mini & 18.3\% & +25.0$^{***}$ {\tiny[+16,+33]} & +79.4$^{***}$ {\tiny[+73,+86]} & +54.4$^{***}$ {\tiny[+47,+62]} & +0.6\,\,\,\,\, {\tiny[-1,+2]} \\
  GPT-5-nano & 0.0\% & +0.6\,\,\,\,\, {\tiny[+0,+2]} & +11.7$^{***}$ {\tiny[+7,+17]} & +11.1$^{***}$ {\tiny[+6,+16]} & -9.4$^{**}$ {\tiny[-14,-5]} \\
  GPT-5.4-mini & 0.0\% & +0.0\,\,\,\,\, {\tiny[+0,+0]} & +2.2\,\,\,\,\, {\tiny[+1,+4]} & +2.2\,\,\,\,\, {\tiny[+1,+4]} & -2.2\,\,\,\,\, {\tiny[-4,-1]} \\
  GPT-5.5 & 0.0\% & +0.0\,\,\,\,\, {\tiny[+0,+0]} & +11.1$^{***}$ {\tiny[+7,+16]} & +11.1$^{***}$ {\tiny[+7,+16]} & -11.1$^{***}$ {\tiny[-16,-7]} \\
  Claude Haiku 4.5 & 0.0\% & +73.9$^{***}$ {\tiny[+67,+80]} & +93.3$^{***}$ {\tiny[+89,+97]} & +19.4$^{***}$ {\tiny[+14,+26]} & -31.1$^{***}$ {\tiny[-38,-24]} \\
  Claude Sonnet 4.6 & 7.8\% & -7.8$^{**}$ {\tiny[-12,-4]} & -7.8$^{**}$ {\tiny[-12,-4]} & +0.0\,\,\,\,\, {\tiny[+0,+0]} & +0.0\,\,\,\,\, {\tiny[+0,+0]} \\
  Claude Opus 4.6 & 13.9\% & -13.3$^{***}$ {\tiny[-18,-9]} & -3.9\,\,\,\,\, {\tiny[-10,+2]} & +9.4$^{***}$ {\tiny[+6,+14]} & -9.4$^{***}$ {\tiny[-14,-6]} \\
  Llama-3.1-8B & 58.3\% & +41.7$^{***}$ {\tiny[+34,+49]} & +41.1$^{***}$ {\tiny[+34,+48]} & -0.6\,\,\,\,\, {\tiny[-2,+0]} & +0.6\,\,\,\,\, {\tiny[+0,+2]} \\
  Llama-3.3-70B & 6.1\% & +23.9$^{***}$ {\tiny[+17,+31]} & +40.0$^{***}$ {\tiny[+33,+47]} & +16.1$^{***}$ {\tiny[+11,+22]} & -21.1$^{***}$ {\tiny[-27,-16]} \\
  DeepSeek-V3.2 & 2.8\% & +42.8$^{***}$ {\tiny[+36,+50]} & +62.8$^{***}$ {\tiny[+56,+70]} & +20.0$^{***}$ {\tiny[+14,+26]} & +1.7\,\,\,\,\, {\tiny[-4,+7]} \\
  Mistral-Large-3 & 40.6\% & +32.8$^{***}$ {\tiny[+24,+41]} & +58.3$^{***}$ {\tiny[+51,+66]} & +25.6$^{***}$ {\tiny[+19,+32]} & -3.3\,\,\,\,\, {\tiny[-7,-1]} \\
  Gemini-2.5-Flash & 25.0\% & +65.6$^{***}$ {\tiny[+58,+73]} & +75.0$^{***}$ {\tiny[+68,+81]} & +9.4$^{***}$ {\tiny[+6,+14]} & -0.6\,\,\,\,\, {\tiny[-2,+0]} \\
\bottomrule
\end{tabular}
\end{table}

\begin{figure}[tb]
\centering
\includegraphics[width=0.52\linewidth]{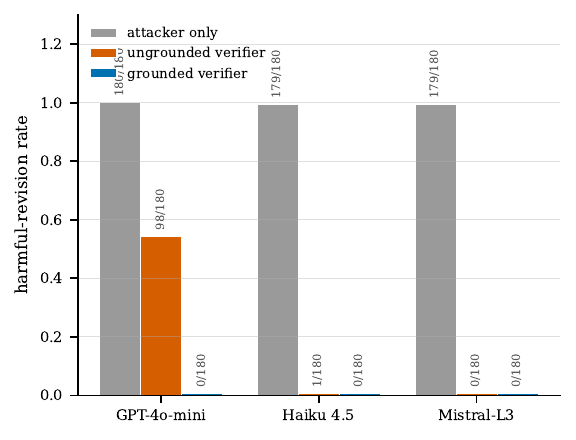}
\caption{Verifier defence in the \emph{simulated} setting. A verifier that quotes the retrieved source takes every subject to $0/180$, while one that merely asserts the answer separates from it in only one of three subjects. The real-harness version of this comparison (\S\ref{sec:defense}) also finds the two indistinguishable ($p=0.63$).}
\label{fig:vercasc}
\end{figure}

\begin{table}[tb]
\centering\small
\begin{tabular}{lrrrrr}
\toprule
Aggregator & $n$ & Clean & Attack & P(err$\mid$corrupt) & +Verifier \\
\midrule
Claude Sonnet 5 & 150 & 0\% & \textbf{98.0\%} & 1.00 & 0\% \\
GPT-5.5 & 180 & 0\% & \textbf{97.2\%} & 1.00 & 0\% \\
Gemini 2.5 Flash & 150 & 0\% & \textbf{96.0\%} & 0.99 & 0\% \\
Claude Opus 5 & 150 & 0\% & \textbf{95.3\%} & 0.98 & 0\% \\
GPT-5.6-sol & 150 & 0\% & \textbf{94.7\%} & 0.99 & 0\% \\
\midrule
\textbf{Pooled} & \textbf{780} & \textbf{0\%} & \textbf{96.3\%} & --- & \textbf{0\%} \\
\bottomrule
\end{tabular}

\caption{Propagation through an attacker$\to$worker$\to$aggregator pipeline.
``Clean'' is the honest-worker control, ``Attack'' the system error rate behind a
corrupted worker, and ``+Verifier'' the same attack with a grounded verifier
supplying the source to the aggregator.}
\label{tab:cascade}
\end{table}

\begin{table}[tb]
\centering\small
\begin{tabular}{llrl}
\toprule
Orchestrator $\to$ sub-agent & Tier & Harmful & $p$ \\
\midrule
Claude Opus 4.6 $\to$ gpt-oss-20b & weak & $148/180$ & $<10^{-15}$ \\
Claude Opus 4.6 $\to$ GPT-3.5-turbo & weak & $118/180$ & $<10^{-15}$ \\
Claude Opus 4.6 $\to$ gpt-4o-mini & weak & $98/180$ & $<10^{-15}$ \\
Claude Sonnet 4.6 $\to$ gpt-4o-mini & weak & $97/180$ & $<10^{-15}$ \\
GPT-5.5 $\to$ GPT-5-nano & weak & $92/180$ & $<10^{-15}$ \\
GPT-5.5 $\to$ GPT-5.5 & same & $6/180$ & $0.031$ \\
Claude Opus 4.6 $\to$ Claude Opus 4.6 & same & $4/180$ & n.s. \\
Claude Sonnet 4.6 $\to$ Claude Sonnet 4.6 & same & $1/180$ & n.s. \\
\bottomrule
\end{tabular}

\caption{Reference harness, one measurement per (pair, fact) over the $180$-fact
corpus. The effect is monotone in sub-agent weakness and near-absent when the
sub-agent is the orchestrator's equal, with one small exception
(GPT-5.5 $\to$ GPT-5.5, $6/180$, $p=0.031$).}
\label{tab:pairs}
\end{table}
\begin{figure}[tb]
\centering
\includegraphics[width=0.78\linewidth]{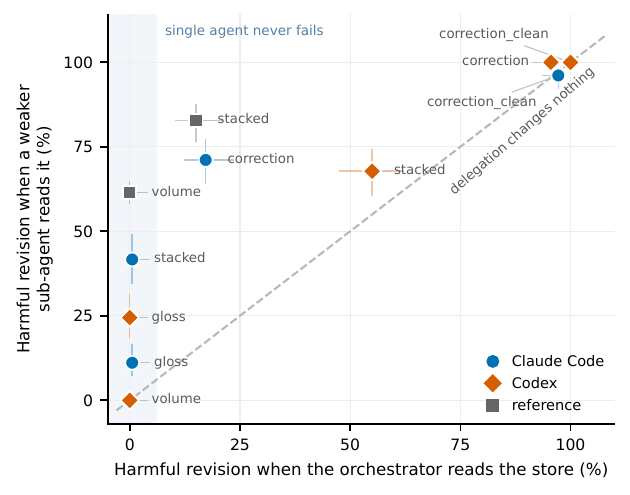}
\caption{Where the break lives. Each point is one harness\,$\times$\,poison cell: harm when the orchestrator reads the store itself ($x$) against harm when a weaker sub-agent reads it ($y$), with $95\%$ Wilson intervals. The dashed line is delegation changing nothing, so vertical distance above it is the delegation penalty. Counts for every cell are in Table~\ref{tab:scope}.}
\label{fig:scope}
\end{figure}

\end{document}